\documentclass[12pt]{iopart}
\usepackage{iopams}
\usepackage{units}
\usepackage{float}
\usepackage{cite}
\usepackage{bm}
\usepackage{color}
\usepackage{graphics}
\usepackage{graphicx}
\usepackage[figtopcap]{subfigure}
\begin{document}

\title[]{Collapse and revival of a Dicke-type coherent narrowing in potassium vapor confined in a nanometric-thin cell}

\author{A	Sargsyan$^1$,  Y Pashayan-Leroy$^2$, C Leroy$^2$ and D Sarkisyan$^1$}

\address{$^1$Institute for Physical Research, NAS of Armenia, Ashtarak, Armenia}
\address{$^2$Laboratoire Interdisciplinaire Carnot de Bourgogne, UMR CNRS 6303, Universit\'{e} de Bourgogne, France}
\ead{davsark@yahoo.com}
\vspace{10pt}
\begin{indented}
\item[]8, November 2015

\end{indented}
\begin{abstract}
A nanometer-thin-cell (in the direction of laser beam propagation) has been elaborated 
with the thickness of the atomic vapor column varying smoothly in the range of $L = \unit[50-1500]{nm}$. 
The cell allows one to study the behavior of the resonance absorption over the $D_1$ line of potassium
atoms by varying the laser intensity and the cell thickness from $L = \lambda / 2$ to $L = 2 \lambda$ with the step $\lambda/2$ 
($\lambda =\unit[770]{nm}$ is the
resonant wavelength of the laser). It is shown that despite the huge Doppler
broadening ($>\unit[0.9]{GHz}$ at the cell temperature $\unit[170]{^{\circ}C}$), at low laser intensities a narrowing of the resonance
absorption spectrum is observed for $L = \lambda/2$ ($\sim \unit[120]{ MHz}$ at FWHM)  and $L = 3/2 \lambda$, whereas 
for $L = \lambda$ and $L =2\lambda$
the spectrum broadens. At moderate laser intensities narrowband velocity selective optical pumping (VSOP)
resonances appear at $L = \lambda$ and $L=2\lambda $ with the linewidth close to the natural one. 
A comparison with saturated
absorption spectra obtained in a 1-cm-sized K cell is presented. 
The developed theoretical model well describes the experiment.
\end{abstract}

%
%
%
%
%
\section{Introduction}
Several experimental realizations and theoretical predictions concerning  thin cells ($\unit[5 -100]{\mu m}$) 
of atomic vapor  have been reported  in the literature~\cite{Izmailov,Vartan,Briaud96,Zambon,Briaud98,Briaud99}.
Recently developed optical nanometer-thin-cell (NTC) containing atomic vapor of alkali metal allows one to 
observe a number of spectacular phenomena, which are absent in common $\unit[0.1-10]{cm}$-long cells, 
particularly: 1) the cooperative Lamb shift caused by dominant contribution of atom-atom interactions~\cite{1SarkPRL2012}; 
2) the largest negative group index measured to date $n_g = -10^5$ caused by propagation of near-resonant 
light through a gas with $L= \lambda/2$ thickness with very high density~\cite{2SarkPRL2012}; 3) strong broadening and shifts of 
resonances, which become significant when thickness $< \unit[100]{ nm}$ caused by atom-surface van der Waals interactions 
due to the tight confinement in the NTC~\cite{Hamdi,SarkEPL2007,SarkPRL2014,SarkPRA2015}.\\
\indent Below is presented another spectacular phenomenon obtained in the NTC filled with potassium vapor. 
Earlier it was shown that an important parameter determining the spectral width, the lineshape and 
the absorption and fluorescence in these cells is the ratio $L/\lambda$ with $L$ being the thickness of the vapor 
column and $\lambda$ the wavelength of the laser radiation resonant with the atomic transition~\cite{Dicke,Duc2003,Duc2004,Duc2007,Cart99,Sark2008}.
Particularly, for Rb and Cs atomic vapors it was demonstrated  that the spectral linewidth of the resonant 
absorption reaches its minimum value at the thickness $L = (2n + 1)\lambda/2$ ($n$ is an integer); this effect has 
been termed the Dicke coherent narrowing (DCN). It was shown that for the thickness $L = n\lambda $ the spectral 
width of the resonant absorption reaches its maximum value close to the Doppler width (of the order of 
several hundreds of MHz) recorded in common cells of conventional length ($\unit[1-10]{cm}$). This 
phenomenon has been termed collapse of DCN (the phenomenon is also absent in common cells)~\cite{Dicke,Duc2003,Duc2004,Duc2007,Cart99,Sark2008}.\\
\indent Recently a one-dimensional NTC filled with natural potassium has been constructed. In order to 
produce K atomic vapor with a density of $\sim \unit[10^{13}]{at/cm^3}$ (it is needed to registrate absorption  
at the  thickness $L=\unit[385]{nm}$) the operation temperature of the NTC's side-arm must be around $\unit[150]{^{\circ}C}$ 
and by $\unit[20]{^{\circ}C}$ higher at the NTC windows (to prevent the condensation of K atoms at the NTC windows). For 
such relatively high temperatures there is a huge Doppler broadening ($> \unit[0.9]{GHz}$). This value is much 
larger than those of Rb ($\unit[0.5]{GHz}$) and Cs ($\unit[0.4]{ GHz}$) atomic vapors~\cite{Demtr}. 
Thus, it was not clear beforehand whether DCN and collapse of DCN will be well observable 
in K vapor. As it is demonstrated below the effect is well observable for potassium vapor, too.	
It  is also shown that the K NTC is a convenient tool for laser spectroscopy.

\section{Experiment}
\subsection{A nano-cell filled with potassium (K) vapor}
A one-dimensional  nano-metric-thin cell filled with natural potassium 
(93.3\% $^{39}$K and 6.7\%  $^{41}$K) has been constructed and used for the experiment. The design of the NTC 
is similar to that of the extremely thin cell described earlier~\cite{Sark2001} (the details of NTC design can be 
found in~\cite{1SarkPRL2012}). The NTC allows one to exploit a variable vapor column thickness $L$ in the range of $\unit[50-1500]{ nm}$. 
The windows of the NTC are fabricated with well-polished crystalline sapphire with the C-axis perpendicular 
to the window surface to minimize birefringence. The NTC operates with a specially designed oven with two ports for laser beam transmission.
\subsection{Experimental setup}
\begin{figure}[hbtp]
\centering
\includegraphics[scale=0.8]{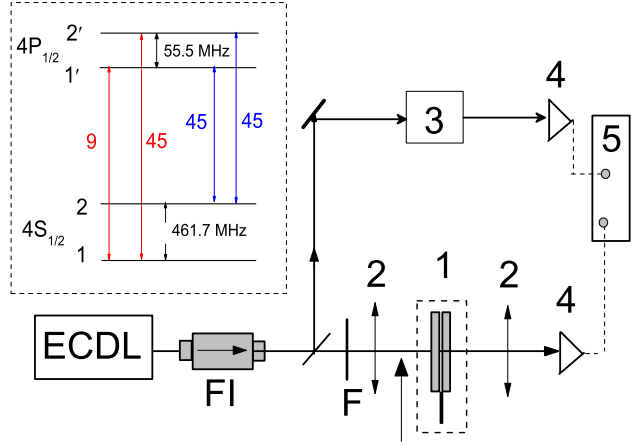}
\caption{Sketch of the experimental setup. ECDL: diode laser; FI: Faraday isolator; 1 - K NTC with a variable thickness;
F-neutral filter, 2 - lenses with a focal length 20 cm: 3 - reference, saturated absorption  scheme;
4 - photodetectors; 5 - oscilloscope. The inset shows the energy levels of the $^{39}$K $D_1$ line.}
\label{fig:SetUp}
\end{figure}
The experimental arrangement is sketched in Fig.~\ref{fig:SetUp}. The linearly polarized beam of an extended cavity 
diode laser (with the linewidth $< \unit[1]{MHz}$), resonant with the $^{39}$K $D_1$ line after passing through 
a Faraday isolator (FI), was focused onto a $\unit[0.5]{mm}$ diameter spot on the K NTC orthogonally 
to the cell window. The transmission signal was detected by a photodiode (4) and was recorded by a 
Tektronix TDS 2014B four-channel storage oscilloscope (5). To record the transmission spectra,
the laser radiation was linearly scanned within up to a $\sim \unit[1]{GHz}$ spectral region covering the studied 
group of transitions. In order to measure the transmission spectra at different NTC thicknesses, 
the oven with the NTC was smoothly moved vertically as indicated by an arrow in Fig.~\ref{fig:SetUp} 
(note that although it is technically easier to move only the ETC, in this case the temperature 
regime of the cell will be changed during the movement). About 30\% of the laser power was 
branched to form the reference spectrum using the saturated absorption (SA) 
scheme (3) with 1.4-cm long potassium cell~\cite{Duc96,Pahwa,Hanley,Hughes}.
\subsection{Experimental results}
\begin{figure}%
\centering
\parbox{2.9in}{%
\includegraphics[scale=0.8]{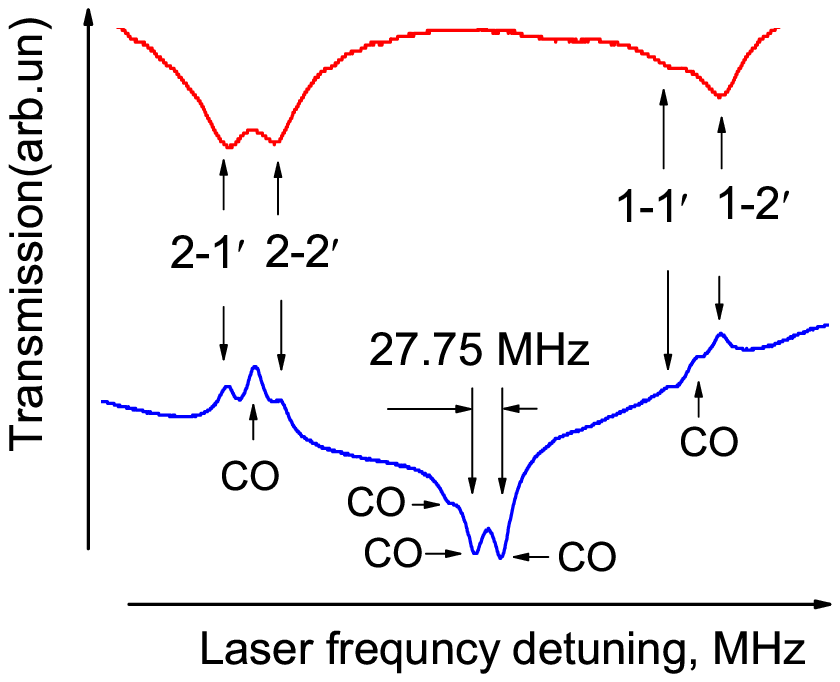}
\caption{Experimental transmission  spectrum (upper curve) of the $F=2,1 \rightarrow F^\prime=1,2$
transitions on the  $^{39}$K $D_1$ line at a cell thickness
$L= \lambda/2 $ and for a low laser power $P_L= \unit[2]{ \mu W}$.
The lower curve is the SA spectrum.
Five cross-over (CO) resonances are marked by the arrows.}%
\label{fig:Fig2}}%
\qquad
\parbox{2.9in}{%
\includegraphics[scale=0.8]{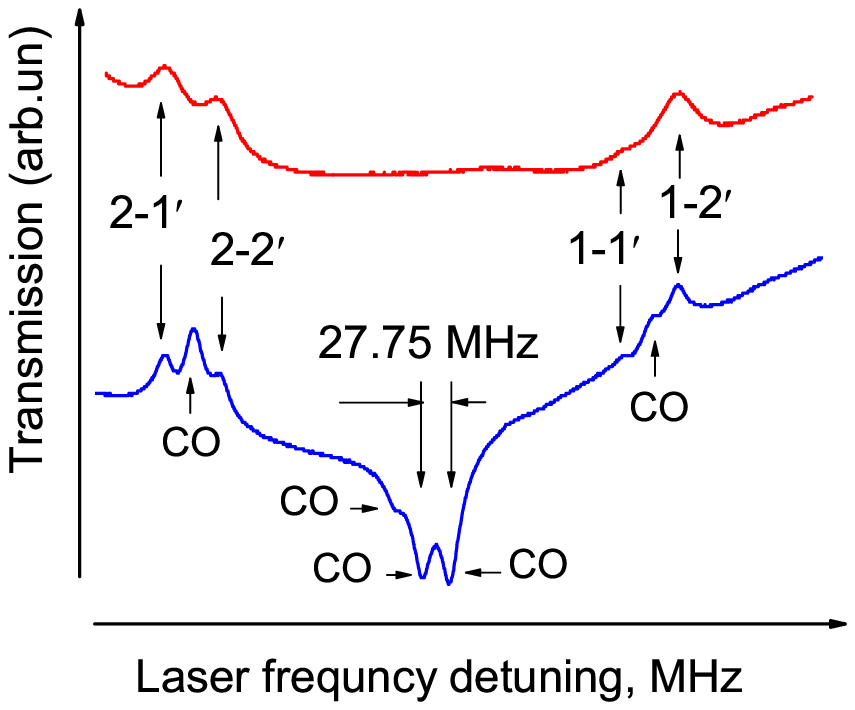}
\caption{Experimental transmission  spectrum (upper curve) of the $F=2,1 \rightarrow F^\prime=1,2$
transitions on the  $^{39}$K $D_1$ line at a cell thickness
$L= \lambda $ and for the laser power $P_L= \unit[60]{ \mu W}$.
The lower curve is the SA spectrum.
Five cross-over (CO) resonances are marked by the arrows.}%
\label{fig:Fig3}}%
\end{figure}%

Consider first the two cases with cell thicknesses $L= \lambda/2$ ($\unit[385]{ nm}$) and $L=  \lambda$
($\unit[770]{ nm}$), the most 
contrasted ones with respect to the length dependence of DCN. 
Figures~\ref{fig:Fig2} and ~\ref{fig:Fig3} show the transmission spectra (the upper curves)
 on the $^{39}$K $D_1$ line for a low laser power
$P_L= \unit[2]{ \mu W}$. The first figure shows the spectrum
at a cell thickness $L=\lambda/2$, while the second one
shows the spectrum for $L=\lambda$.
As  
seen from Fig.~\ref{fig:Fig2} there is Dicke-narrowing and all of the four atomic transitions $F=2, 1 \rightarrow F^\prime=1, 2$ 
have sub-Doppler spectral linewidths ($\sim \unit[120]{MHz}$ at FWHM, while the 
low- and the high-frequency wings of the lineshape extend to a few hundreds of MHz) and are 
partially resolved. Note, that there is a narrowing by nearly 8 orders of magnitude 
in the transition linewidths with respect to
the Doppler width and the line profile is similar to the profile 
of time-of-flight broadening~\cite{Meschede}.\\
\indent For $L= \lambda/2$, the Dicke-narrowed sub-Doppler absorption profiles simply broaden and 
saturate in amplitude when increasing the light intensity. Meanwhile for the thickness $L = \lambda$ 
the spectral width of the resonant absorption spectrum reaches its maximum value (it means the
 collapse of DCN) close to the Doppler width (of the order of 1 GHz) and there appear
 sub-Doppler dips of reduced absorption at the line-center on the broad absorption profile (see Fig.~\ref{fig:Fig3})
 with the  increase of light intensity. These narrowband features 
 resulting from the hyperfine pumping between the two hyperfine components of 
 the ground $4S_{1/2}$ state via the upper excited hyperfine
states  are called
 velocity selective optical pumping (VSOP) resonances, that are located at four $F=1,2 \rightarrow F^\prime= 1, 2$   transitions. 
 The linewidth of the VSOP could be close to the natural one ($\sim \unit[6]{MHz}$)~\cite{Duc2007,Ghosh}.\\
 \indent It should be noted, that the use of NTC as the frequency reference for the $^{39}$K atomic transitions 
 in comparison with the use of the SA spectra (see Fig.~\ref{fig:Fig3}) provides several advantages which 
 are as follows : (i) 
 contrary to SA, in the case of NTC, the crossover lines are absent, and this is very important
  to study the Zeeman spectra; (ii) for the case of NTC where the thickness $L = \lambda$, the ratio
   of the amplitudes for the VSOP peaks is close to the ratio of the atomic probabilities 
   of the corresponding transitions (for example, as seen from the inset of Fig.~\ref{fig:SetUp} the 
    atomic transitions $F=2 \rightarrow F^\prime=1$  and $F=2 \rightarrow F^\prime=2$  
   have the same probabilities which is confirmed by 
   the upper curve in Fig.~\ref{fig:Fig3}), while it is not the case for SA; moreover, 
   as a rule, a strong cycling transition in the SA spectrum has a smaller amplitude in 
   comparison with that of the weaker non-cycling one; (iii) the SA geometry requires 
   counter propagating beams, while a single-beam transmission spectrum is needed for 
   a NTC; and (iv) the laser power needed for spectral reference applications in  
   case of NTC is more than by an order of magnitude less than that needed for SA. 
   The above mentioned advantages demonstrated earlier for the 
   frequency reference based on a Rb NTC~\cite{Moi} are also valid for a K NTC.\\
\begin{figure}
\begin{center}
\subfigure[]{
\resizebox*{6.6cm}{!}{\includegraphics{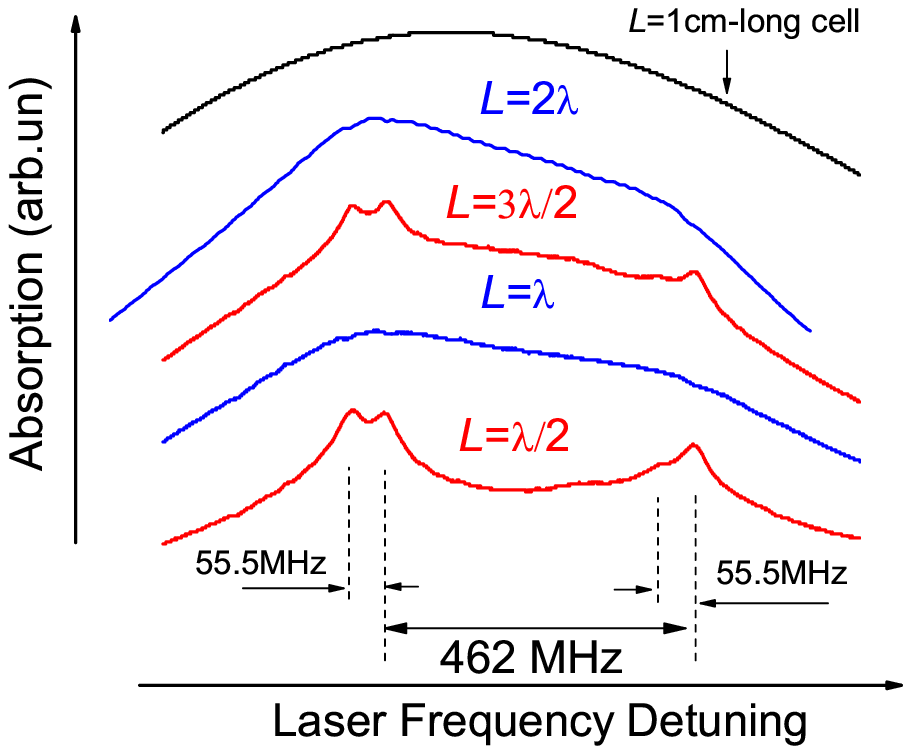}
} \label{fig:Fig4a}
}\hspace{6pt}
\subfigure[]{
\resizebox*{6.8cm}{!}{\includegraphics{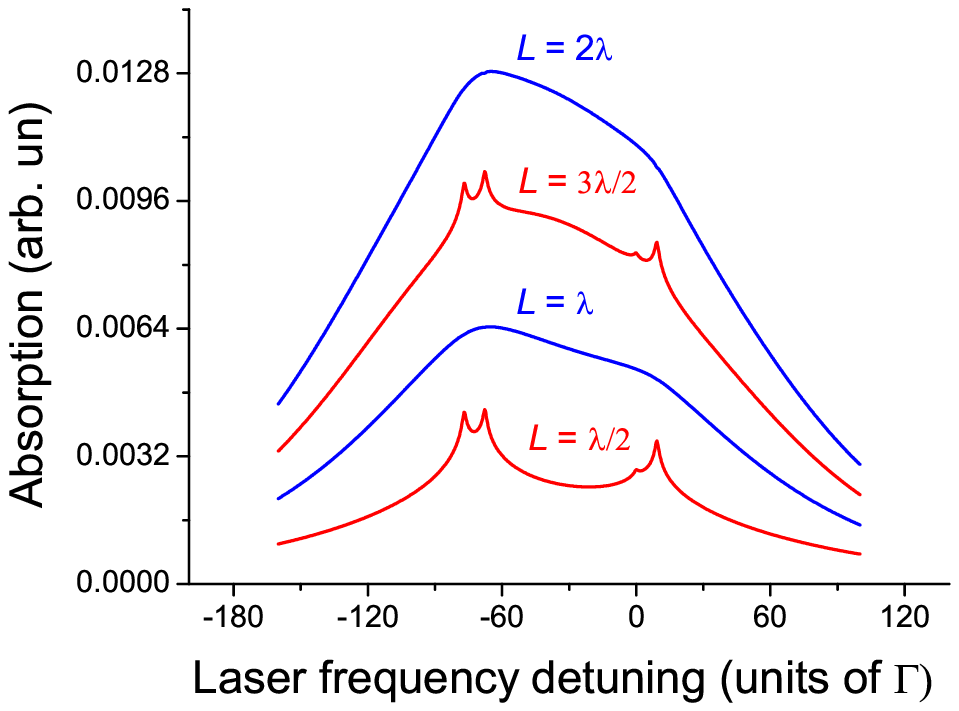}
} \label{fig:Fig4b}
}
\caption{a) Experimental absorption spectra for the atomic 
transitions $F=2,1 \rightarrow F^\prime=1,2$ for the low laser power $P_L= \unit[2]{\mu W}$ at
different NTC thicknesses: $L= \lambda/2$, $L=\lambda$, $L= 3\lambda /2$ and $L= 2\lambda$. The side-arm temperature
of the NTC is $\unit[150]{^{\circ}C}$. The upper curve shows the absorption in a common 1.4-cm-sized K cell
(cell temperature $\unit[55]{^{\circ}C}$). b) Theoretical absorption spectra for a low laser power with the
 Rabi frequency  $\Omega =\unit[0.05]{MHz}$ for different NTC
thicknesses: $L= \lambda/2$, $\lambda$, $3 \lambda/2 $ and $2\lambda$. }
\label{fig:Fig4}
\end{center}
\end{figure}
   \indent The experimental absorption spectra for the atomic transitions $F=2,1\rightarrow F^\prime=1,2$  for the low 
   laser power $P_L= \unit[2]{ \mu W}$ are shown in Fig.~\ref{fig:Fig4}\subref{fig:Fig4a} for different NTC thicknesses 
   varying from  $L= \lambda/2$ to $L= 2\lambda$  with the step
   of $ \lambda/2$. As we see for $L= \lambda/2 $ and $L= 3 \lambda/2$, there is Dicke-narrowing that's why all of the  
   four atomic transitions $F=1,2 \rightarrow F^\prime= 1, 2$  have sub-Doppler spectral linewidths and are partially 
   resolved (the best spectral resolution is achieved for $L= \lambda/2$). Meanwhile for the thicknesses 
   $L = \lambda$ and $2 \lambda $ the spectral widths of the resonant absorption spectra reach their maximum 
   values (it means the collapse of the Dicke-narrowing). 
\begin{figure}
\begin{center}
\subfigure[]{
 \resizebox{0.37\textwidth}{!}{\includegraphics{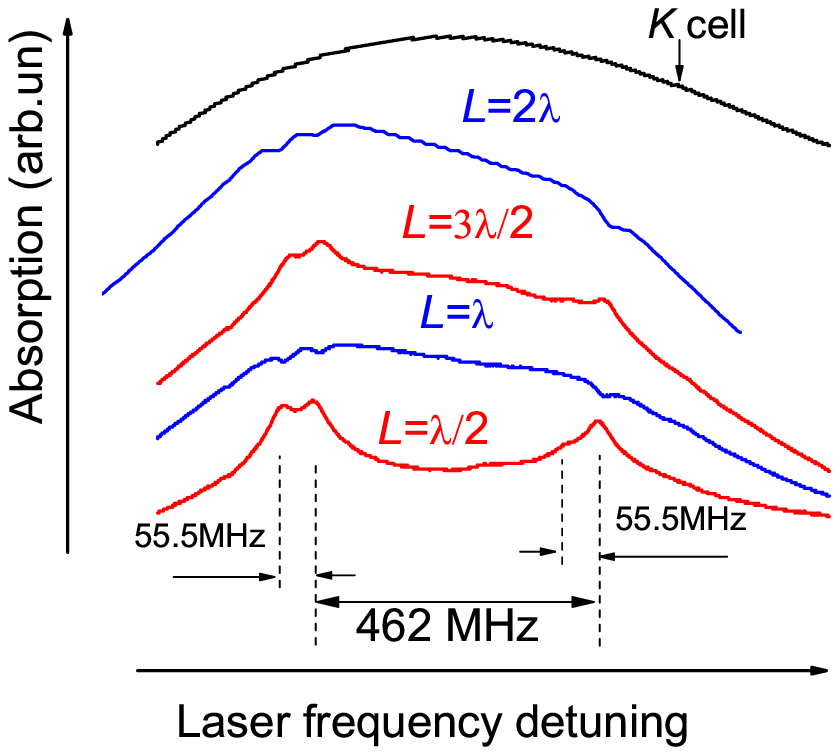}
 }
 \label{fig:Fig5a}
} \vspace{0.005cm}
 \subfigure[]{
 \resizebox{0.41\textwidth}{!}{\includegraphics{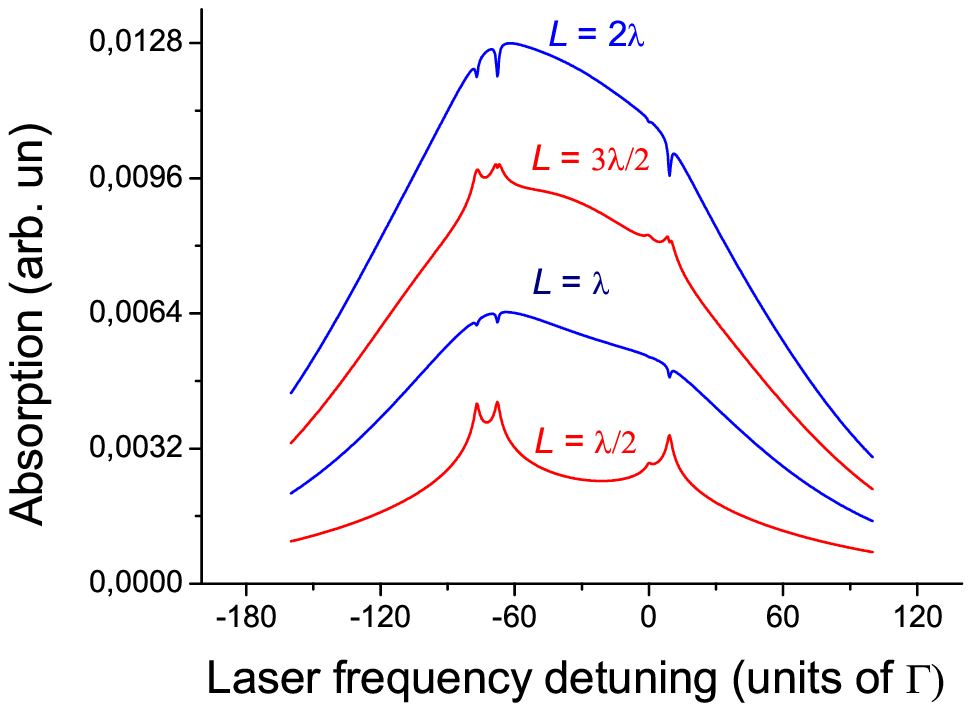}
 }
 \label{fig:Fig5b} 
 }
\caption{a)  Experimental absorption spectra for the atomic
transitions $F=2,1 \rightarrow F^\prime=1,2$ for a moderate laser power, $P_L= \unit[60]{\mu W}$, 
at different NTC thicknesses: 
$L= \lambda/2$, $\lambda$, $3 \lambda/2 $ and $2\lambda$. The upper curve shows the absorption of 1.4-cm-long K cell. 
b) Theoretical absorption spectra for a moderate power with the Rabi frequency of $\Omega =\unit[0.3]{MHz}$  
for different NTC thicknesses: $L= \lambda/2$, $\lambda$, $3 \lambda/2 $ and $2\lambda$. }
\label{fig:Fig5}
\end{center}
\end{figure}
   The theoretical spectra presented in Fig.~\ref{fig:Fig4}\subref{fig:Fig4b} well describe the experiment.\\
\indent Figure~\ref{fig:Fig5}\subref{fig:Fig5a} shows the experimental absorption spectra at the moderate laser power $P_L=\unit[60]{\mu W}$ for the cell thickness $L$ varying from $L= \lambda/2$
to $L= 2 \lambda$ 
with the step of  $\lambda/2$. 
As we see for $L= \lambda/2$ and $L= 3 \lambda/2$, there is still 
Dicke-narrowing (while a little spectrally broadened) that's why the atomic transitions $F=1,2 \rightarrow F^\prime= 1, 2$  
are still partially resolved. Meanwhile for the thicknesses $L = \lambda$ and $2 \lambda$ the spectral widths
 of the resonant absorption spectra reach their maximum values and there appear small narrowband VSOPs. 
 The calculated absorption spectra (see Fig.~\ref{fig:Fig5}\subref{fig:Fig5b}) well describe the experiment.\\
 \begin{figure}
\begin{center}
\subfigure[]{
\resizebox*{0.37\textwidth}{!}{\includegraphics{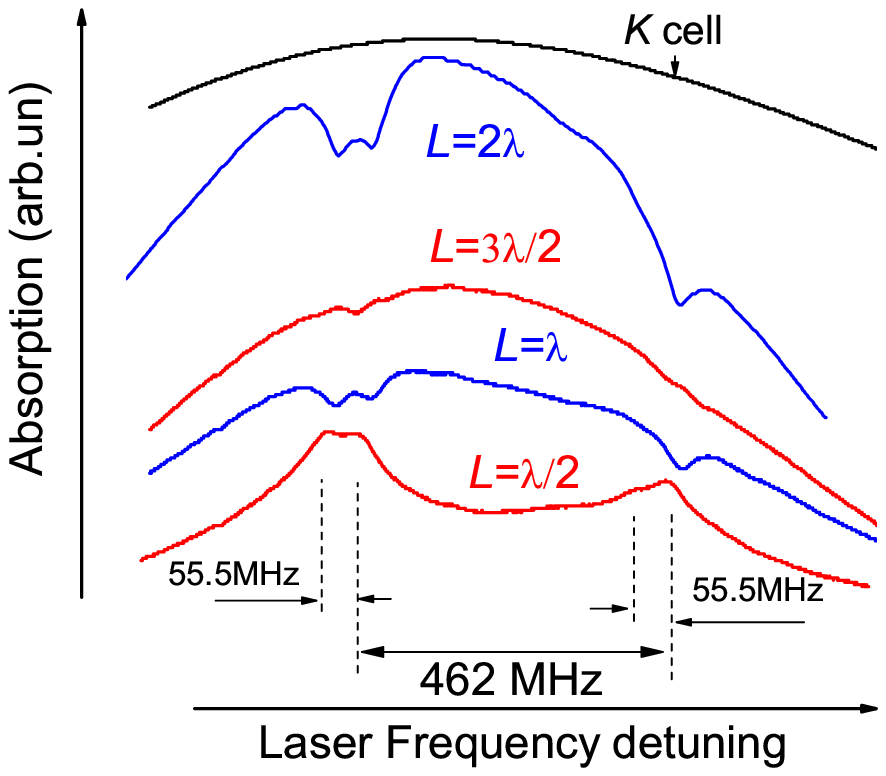}
}\label{fig:Fig6a}
}\hspace{6pt}
\subfigure[]{
\resizebox*{6.0cm}{!}{\includegraphics{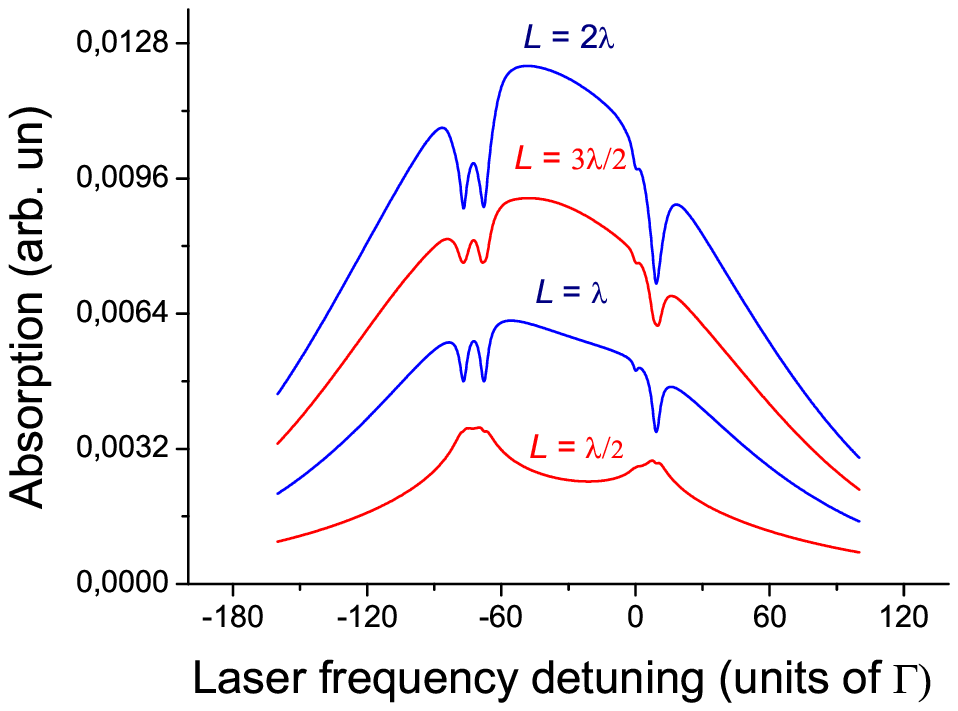}
}\label{fig:Fig6b}
}
\caption{a) Experimental absorption spectra for the atomic
transitions $F=2,1 \rightarrow F^\prime=1,2$ for a high laser power, $P_L= \unit[2.5]{ mW}$
at different NTC thicknesses: $L= \lambda/2$, $\lambda$, $3 \lambda/2 $ and $2\lambda$.
The upper curve shows the absorption of 1.4-cm-long K cell.
b) Theoretical absorption spectra for a high laser power  with the Rabi frequency of $\Omega =\unit[2]{MHz}$
for different NTC thicknesses: $L= \lambda/2$, $\lambda$,
$3 \lambda/2$ and $2\lambda$.}
\label{fig:Fig6}
\end{center}
\end{figure}
   \indent The experimental absorption spectra for the high laser power of $P_L=\unit[2.5]{ mW}$ 
   for the cell thickness $L$ varying from $L= \lambda/2$
to $L= 2 \lambda$
with the step of  $\lambda/2$  are shown in Fig.~\ref{fig:Fig6}\subref{fig:Fig6a}. 
As we see for $L= \lambda/2$ and $L= 3 \lambda/2$, there is still Dicke-narrowing while 
spectrally broadened that's why the atomic transitions $F=1,2 \rightarrow F^\prime= 1, 2$
   are worst resolved. Meanwhile for the thicknesses $L = \lambda$ and $2 \lambda$ the spectral widths of 
   the resonant absorption spectra reach their maximum values and there appear large narrowband VSOPs. 
   Note that at high laser intensities VSOP resonances appear also for $L = 3/2\lambda$.
   The theoretical spectra presented in Fig.~\ref{fig:Fig6}\subref{fig:Fig6b} well describe the experiment.\\
   \indent As it was demonstrated earlier for Rb vapor~\cite{Duc2007} and is confirmed for 
    K vapor (see Figs.4 (a), (b) and Figs.~\ref{fig:Fig6}\subref{fig:Fig6a},~\ref{fig:Fig6}\subref{fig:Fig6b}) the peak absorption at $L= \lambda/2$
   is only slightly smaller than that of $L=\lambda$  at low intensities (the absolute value 
   of the peak absorption is $\sim 0.6 \%$ for $\unit[150]{^{\circ}C}$ at the side-arm), and the absorption 
   for $L= \lambda/2$ becomes even larger than that of $L=\lambda$  at higher intensities.\\
   \indent With further increasing of the NTC thickness $L$, the Dicke-narrowed sub-Doppler absorption 
   profiles simply broaden, their relative contrast decrease and the oscillating character of the
    Dicke narrowing with the cell thickness vanishes (for Cs vapor the narrowing completely 
    vanishes for $L > \unit[5]{\mu m}$~\cite{Briaud98}). As to the VSOPs as it was demonstrated in~\cite{Briaud99} they are still 
    observable in Cs vapor at $L \sim \unit[10]{\mu m}$
    (although with a very small amplitude and small signal/noise ratio).\\
    \indent It should be noted, that the presence of the narrow-band
    resonances at $L= \lambda/2$ (absorption) and $L= \lambda$ (VSOPs)
    in the NTC filled with $^{39}$K allows us to detect complete hyperfine
    Paschen-Back regime at relatively small magnetic fields~\cite{Ler2015}.
\section{Theoretical model}
\begin{figure}[hbtp]
\centering
\includegraphics[scale=1.0]{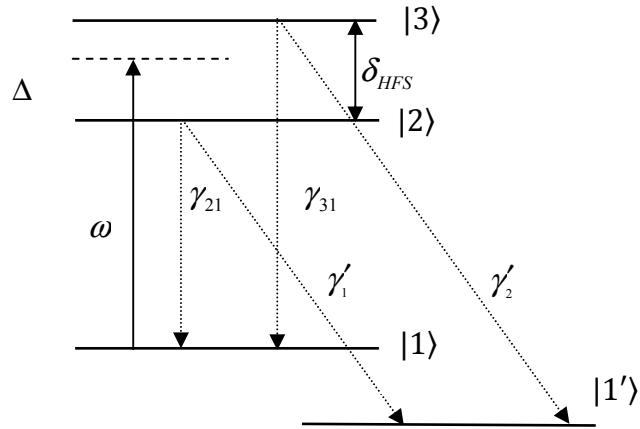}
\caption{The three-level scheme around the $D_1$ line of $^{39}$K as described in the text.}
\label{fig:Model}
\end{figure}

We report here the theoretical
model aimed at reproducing the features of absorption
spectra registered in the experiment presented in this paper.
The model is close to the one described in~\cite{Malak}.
We
consider a three-level model with the states assigned to
the real states of the
$D_1$ line $4S_{1/2} \rightarrow 4P_{1/2}$ of potassium atom (see the inset in Fig.~\ref{fig:SetUp}).
The scheme of the light-atom interaction  is presented in Fig.~\ref{fig:Model}.
As a ground state $|1 \rangle$ we use one of the hyperfine components
of the $4S_{1/2}$ ground level  ($F=1$ or $F=2$) which
is coupled to the excited states $|2 \rangle$ ($ F^\prime=1 $) and
 $|3 \rangle$ ( $F^\prime=2 $)
by a laser field of amplitude $E$ and frequency $\omega$.
The state denoted by $|1^\prime\rangle$ corresponds to the other ground hyperfine component
not taken into the model.
The electromagnetic field is assumed to be a plane wave propagating
through a gas of K atoms confined in a cell of length $L$
along $z$-direction.
Our consideration is conducted
within the semiclassical approximation which means the light field is treated classically
while the atomic medium is quantized.
The detuning of the laser field from the transition
$|1 \rangle \rightarrow | 2\rangle$
 is denoted by
 $\Delta=\omega_{21}-\omega - kv_z$, where $kv_z$ is the Doppler shift contribution for the
detuning of the laser field corresponding to the velocity
component $v_z$ along the propagation vector $\textbf{k}$. \\
\indent The total Hamiltonian in the electric-dipole and rotating-wave
approximations (RWA)~\cite{Shore} for atom-field interacting system can be written as
\begin{equation}
H = H_0+H_I.
\end{equation}
$H_0$ is the unpertubed (i.e. bare) three-state atomic Hamiltonian
\begin{equation}
H_0 = \hbar\Delta|2\rangle \langle 2| + \hbar (\Delta + \delta_{\rm {HFS}}) |3\rangle \langle 3|
\end{equation}
with
$\delta_{\rm {HFS}}$ being the hyperfine splitting between the upper excited levels,
and $H_I$ the atom-field interaction Hamiltonian
\begin{equation}
H_I = -\hbar \Omega_{1}^*|1\rangle \langle 2| - \hbar \Omega_{2}^* |1\rangle \langle 3|  + h.c.
\end{equation}
Here
$\Omega_{1,2}=d_{1i}E/2\hbar$  are the Rabi frequencies
of the corresponding atomic transitions, with $d_{1i}$ ($i=2,3$)  being the electric-dipole matrix element
associated with the transition $|1\rangle \rightarrow |i\rangle$, and $h.c.$
represents Hermitian conjugate.\\
\indent We point out the key points of the basic assumptions
made in the model: the atomic number
density is assumed to be low enough so that the effect of
collisions between the atoms can be ignored and
only atom-surface collisions are to be considered; the atoms experience
inelastic collisions with the cell walls; the incident
beam diameters largely exceed the cell thickness so the decoherence due to the
diffusion of atoms out of the incident beams is neglected.
	The effect of the reflection of the radiation from
the highly-parallel windows of the nano-cell behaving as a Fabry-Perot cavity~\cite{DucJOSAB} is taken
into account.  The Doppler-broadening effect is included assuming a Maxwellian
distribution of velocities.\\
\indent The internal atomic dynamics is described by a semiclassical density matrix
$\rho=\sum \rho_{ij} |i\rangle \langle j|$, the time evolution of which is given by the Liouville
equation of motion~:
\begin{equation}
\dot{\rho}=-\frac{i}{\hbar}[\hat{H},\rho]+\Gamma \rho,
\end{equation}
where $\Gamma$ is the relaxation operator. The density-matrix equations of motion for a considered three-level system read
\renewcommand{\Im}{\mathrm{Im}}
\renewcommand{\Re}{\mathrm{Re}}
\begin{equation}
\eqalign{\rho_{11} = -2\,\Im(\Omega_1^* \rho_{21}) - 2\,\Im(\Omega_2^* \rho_{31}) + \gamma_{21} \rho_{22} + \gamma_{31} \rho_{33}, \cr 
\rho_{22} =  2\,\Im(\Omega_1^* \rho_{21}) - 2\Gamma \rho_{22}, \cr
\rho_{33} =  2\,\Im(\Omega_2^* \rho_{31})  - 2\Gamma \rho_{33}, \cr 
\rho_{21} =  i \Omega_1 (\rho_{11} - \rho_{22}) - i \Omega_2\rho_{32}^* - (\Gamma  + i \Delta) \rho_{21}, \cr 
\rho_{31} =  i \Omega_2 (\rho_{11} - \rho_{33}) - i \Omega_1 \rho_{32} - \left [\Gamma  + i (\Delta + \delta_{HFS})\right ] \rho_{31}, \cr 
\rho_{32} =  i \Omega_2 \rho_{21}^* - i \Omega_1^* \rho_{31} - (2\Gamma +  i \delta_{HFS}) \rho_{32}, }
\label{eq:DensityMatrix}
\end{equation}
and $\rho_{ij}=\rho_{ji}^*$. Here $\gamma_{21}$ and $\gamma_{31}$ are the spontaneous decay
rates from the excited states $|2\rangle$ and $|3\rangle$ to the ground state $|1\rangle$, 
$2\Gamma=\gamma_N=\gamma_{21} + \gamma_1^\prime=\gamma_{31} + \gamma_2^\prime$
is the total rate of the spontaneous decay with
$\gamma_1^\prime$, $\gamma_2^\prime$ being the decay rates describing the optical hyperfine pumping
 between the two hyperfine components
of the ground $4S_{1/2}$ state. For our calculations we take 
$\gamma_{21}=1/6\gamma_N$, $\gamma_1^\prime=5/6\gamma_N$ and $\gamma_{31}=0.5\gamma_N$, $\gamma_2^\prime=0.5\gamma_N$,
with the natural linewidth $\gamma_N = \unit[2\pi \times 6.03]{MHz}$ for $^{39}$K.\\
\indent We are concerned with the laser field transmitted through the second window of the cell containing the atomic vapor
\begin{equation}
E_{out} = t_2 E_{0}.
\end{equation}
Here $t_2$ is the transmission coefficient of the second window of the cell, and  $E_{0}(z)$
is the laser field inside the cell.
 The goal is to find an expression for the field $E_{0}(z)$ at any position
 $z$ inside the vapour  in order to obtain the transmitted signal $I_ \propto |E_{out}|^2$.
The field can be represented as
\begin{equation}
E_{0}(z)=E_{0}^{\prime}(z) + E_{0}^{\prime\prime}(z).
\end{equation}
Here $E_{0}^{\prime}(z)$ is the laser field inside the empty cell,
and $E_{0}^{\prime\prime}(z)$ is the resonant contribution of the medium to the field.
Note that for a very dilute medium $E_{0}^{\prime\prime}(z) \ll E_{0}^{\prime}(z)$.
 Under this condition, considering the effects of interference and multiple reflections
from the cell windows one obtains for the laser field
inside the empty cell~\cite{DucJOSAB}
\begin{equation}
\label{eq:EmptyFP}
E_{0}^{\prime}(z) = \mathcal{E}t_1\frac{ 1-r_2 \exp[2ik (L-z)]}{F},
\end{equation}
and for the resonant contribution at the cell exit $z=L$~\cite{Malak}
\begin{equation}
\label{eq:Res}
E_{0}^{\prime\prime}(z=L) = \frac{2\pi ik}{F}\int_0^{z'=L} \mathcal{P}_0(z')\left [ 1 - r_1e^{2ik z'} \right ]dz',
\end{equation}
where $\mathcal{E}$ is the amplitude of the incident laser field and
$\mathcal{P}_0(z)$ is  the amplitude of the polarization  induced inside the
vapor  at the laser frequency. Here the factor $F=1-r_1r_2\exp(2ik L)$ takes into account
the influence of the  beam reflected by the second
wall of the cell (which is important in the case where
the cell thickness is $L\sim\lambda$), i.e. the Fabry-Perot effect, and $r_1$ and $r_2$
are the reflection coefficients of the first and second windows, respectively, and
$t_{1}$ is the transmission coefficient of
the first window.\\ 
Taking into account the dilute character of the medium ($E_{0}^{\prime\prime} \ll E_{0}^{\prime}$),
we obtain for the transmitted signal:
\begin{equation}
I \approx t_2^2 \left [ \left | E_{0}^{\prime}(L) \right |^2 +
2\Re \left \{ E_{0}^{\prime\prime}(z=L)E_{0}^{\prime *}(z=L)\right \}  \right ]  .
\end{equation}
The absorption signal of the laser field is given by
the second term in the brackets
\begin{equation}
\label{eq:AbsSignal}
S_{abs} \sim 2\Re \left \{ E_{0}^{\prime\prime}(z=L)E_{0}^{\prime *}(z=L) \right \} .
\end{equation}
\indent To obtain the absorption spectra, we need to calculate
the polarization of the medium on the frequency of the
laser field that is determined by
\begin{equation}
\label{eq:ProbePolariz}
P_0(z)= N  d_{21} \left ( \rho_{21}^{+} + \rho_{21}^{-} \right ) + N  d_{31} \left ( \rho_{31}^{+} +
\rho_{31}^{-} \right ),
\end{equation}
 where nondiagonal matrix elements $\rho_{j1}^{+} \equiv \rho_{j1}$ ($z=v_zt$)
 and $\rho_{j1}^{-} \equiv \rho_{j1}$ ($z=L-v_zt$) with $j=2,3$ relate to the atoms flying with the velocity
in the positive and negative directions of the cell
axis, respectively. We provide this by exact
numerical solution of Eqs.~(\ref{eq:DensityMatrix})
 by assuming realistic parameters corresponding to the conditions of
our modeled experiment.
To include the Doppler-broadening effect, we average
the density-matrix elements obtained for a single atom
over the velocity distribution of the atoms
which is considered to be Maxwellian
and is given by
$W(v)=\frac{N}{u \sqrt{\pi}}\exp[-(\frac{v}{u})^2]$, where $v$
 is the atomic velocity, $N$ is the atomic
density, and $u$ is the most probable velocity given by
$u = 2k_B T/M$ with $k_B$ being the Boltzmann constant,  $T$  the temperature
of the vapour in Kelvin and
$M$ the atomic mass.
  Inserting
formulas~(\ref{eq:EmptyFP}), (\ref{eq:Res}), (\ref{eq:ProbePolariz}) to
Eq.~(\ref{eq:AbsSignal}) one gets for the Doppler broadened absorption
profile~\cite{Malak}
\begin{equation}
\label{Abs}
\eqalign{ \frac{1}{\mathcal{E}^2} \langle S_{abs} \rangle   = & - \frac{4\pi \omega N t_2^2 t_1}{c u \sqrt{u}}
\frac{1}{\mathcal{E} |F|^2} \int_0^{\infty}e^{-v_z^2/u^2} v_zdv_z \int_0^{L/v}dt\\
 & \times \Im \bigg \{ \sum_{i=2,3}d_{i1} \bigg [  \rho_{i1}^{+} \left( t,\Delta^+,E_{0}(v_z t) \right)
 (1 - r_1 e^{2 i k v_z t})    \\
 &   + \rho_{i1}^{+}\left ( t,\Delta^-,E_{0}(L - v_z t) \right )
 \left (1 - r_1 e^{2 i k ( L - v_z t)} \right) \bigg ]  \bigg \}.}
\end{equation}
Here $\Delta^{+}=\Delta^{+} \pm k v_z$ are the Doppler shifted
detunings of the probe field. The plus
sign refers to the atoms with the velocity $\textbf{v}$ in the positive
direction of the cell axis, and the minus sign refers to the
atoms with the velocity $\textbf{v}$ in the negative direction of the
cell axis.\\
\indent Figures~\ref{fig:Fig4}\subref{fig:Fig4b}-\ref{fig:Fig6}\subref{fig:Fig6b}  present
 the results for the absorption signal as given by the presented theoretical model. 
 Shown are the absorption spectra versus the laser frequency detuning.
 To match the
theoretical spectra to the relevant experimental ones,
a value of the Rabi frequency $\Omega$, was attributed to all the spectra
registered at a given intensity $I$. These values of the Rabi frequency were
estimated from $\Omega/2\pi = a \gamma_N \sqrt{I / 8}$,
where $I$ is the laser intensity in mW/cm$^2$, $\gamma_N/2\pi$  is the decay rate of the excited state (6.03 MHz),
and $a$ is the fit parameter. The fit parameter  $a$ was optimized to
obtain the best match of theoretical and experimental 
features, with regard to the whole series of experimental
spectra, such as presented in Figs.~\ref{fig:Fig4}-\ref{fig:Fig6} (for our case $a  \approx 0.03$). 
The results of our calculations are in general agreement with those of the experiment.

\section{Conclusion}
It is demonstrated both experimentally and theoretically that the spectral linewidth 
of the resonant absorption of potassium $^{39}$K vapor contained in a nano-metric thin
 cell with a variable thickness   reaches its minimum value ($\sim \unit[120]{ MHz}$) at the 
 thicknesses $L = \lambda/2$ (385 nm) and $L = 3\lambda/2$ ($\unit[1155]{nm}$) 
 (despite the huge Doppler broadening of $\sim \unit[1]{GHz}$ at
 the cell temperature of $\sim \unit[170]{^{\circ}C}$). The effect is  termed the Dicke 
 coherent narrowing, earlier observed for  Rb and Cs vapors. For the thicknesses 
 $L = \lambda$ ($\unit[770]{nm}$) and $L = 2\lambda$ ($\unit[1540]{nm}$) the spectral width of the resonant absorption 
 reaches its maximum value close to the Doppler width, recorded in  common 
 cells of conventional length ($\unit[1-10]{cm}$).\\
 \indent For $L= \lambda/2$ and $L=3 \lambda/2$, the Dicke-narrowed sub-Doppler absorption profiles simply 
 broaden and saturate in amplitude when increasing the light intensity. Meanwhile for 
 the thicknesses $L = \lambda$ and $2\lambda$, with the increase in the light intensity there appear narrowband sub-Doppler dips of 
 reduced absorption (VSOPs) at the line-center on the broad absorption profile. 
 Practical applications of the potassium NTC are noted. The calculated theoretical spectra 
 well describe the experiment. A comparison of the NTC spectra with the well known saturated  
 absorption spectra is provided, and the advantages of the NTC use are shown.
 \section*{Acknowledgements }
 The research was conducted in the scope of the International Associated 
 Laboratory IRMAS (CNRS-France \& SCS-Armenia).


\section*{References}
\begin{thebibliography}{0}
\bibitem{Izmailov} Izmailov A Ch 1992 \LP {\bf 2} 762.
\bibitem{Vartan}Vartanyan T A and Lin D L 1995 \PR A {\bf 51} 1959.
\bibitem{Briaud96} Briaudeau S,  Bloch D and Ducloy M 1996 \EPL {\bf 35} 337.
\bibitem{Zambon} Zambon B, Nienhuis G, 1997 {\it Opt.Commun.} {\bf 143}  308.
\bibitem{Briaud98} Briaudeau S, Saltiel S, Nihenius G, Bloch D and Ducloy M 1998 \PR A {\bf 57} R3169.
\bibitem{Briaud99} Briaudeau S,  Bloch D, and Ducloy M 1999 \PR A {\bf 59} 3723.
\bibitem{1SarkPRL2012} Keaveney J, Sargsyan A, Krohn U, Sarkisyan D, Hughes I G and Adams C S 2012 \PRL {\bf 108} 173601.
\bibitem{2SarkPRL2012} Keaveney J, Hughes I G, Sargsyan A, Sarkisyan D and Adams C S 2012 \PRL {\bf 109} 233001. 
\bibitem{Hamdi} Hamdi I \etal 2005 \LP {\bf 15} 987.
\bibitem{SarkEPL2007} Fichet M \etal 2007 {\it Europhys. Lett.} {\bf 77} 54001.
\bibitem{SarkPRL2014}  Whittaker K A, Keaveney J, Hughes I G, Sargsyan A, Sarkisyan D and Adams C S 2014 \PRL {\bf 112} 253201.
\bibitem{SarkPRA2015}  Whittaker K A, Keaveney J, Hughes I G, Sargsyan A, Sarkisyan D and Adams C S \PR A {\bf 92} 052706.
\bibitem{Dicke}  Romer RH and Dicke RH 1955 \PR A {\bf 99} 532.
\bibitem{Duc2003}  Dutier G, Yarovitski A, Saltiel S, Papoyan A, Sarkisyan D, Bloch D and Ducloy M 2003 {\it Europhys. Lett.} {\bf 63} 35.
\bibitem{Duc2004}  Sarkisyan D, Varzhapetyan T, Sarkisyan A, Malakyan Yu, Papoyan A, Lezama A, Bloch D and Ducloy M 2004  \PR A {\bf 69} 065802.
\bibitem{Duc2007}  Andreeva C, Cartaleva S, Petrov L, Saltiel S, Sarkisyan D, Varzhapetyan T, Bloch D and Ducloy M 
2007 \PR A {\bf 76} 013837.
\bibitem{Cart99}  Cartaleva S,  Saltiel S, Sargsyan A, Sarkisyan D, Slavov D, Todorov P, Vaseva K 2009 \JOSA B {\bf 26} 1999.
\bibitem{Sark2008} Varzhapetyan T, Nersisyan A, Babushkin V, Sarkisyan D, Vdovi\'{c} S, Pichler G 2008.
       \jpb  {\bf 41} 185004
\bibitem{Demtr}  Demtroder W 2002 {\it Laser Spectroscopy: Basic Concepts and Instrumentation} (Springer).
\bibitem{Sark2001} Sarkisyan D, Bloch D, Papoyan A, and Ducloy M 2001 {\it Opt. Commun.}  {\bf 200} 201.
\bibitem{Duc96} Bloch D, Ducloy M, Senkov N, Velichansky V and Yudin V 1996 \LP {\bf 6}  670.
\bibitem{Pahwa} Pahwa K, Mudarikwa L, Goldwin J 2012 {\it Opt. Express} {\bf 20} 17456.
\bibitem{Hanley} Hanley R K, Gregory P D, Hughes I G, Cornish S L 2015 \jpb {\bf 48} 195004.
\bibitem{Hughes} Zentile M A, Keaveney J, Weller L, Whiting D J, Adams C S, Hughes I G 2015 {\it Comput. Phys. Commun.} {\bf 189} 162.
\bibitem{Meschede} Meschede Dieter 2007 {\it Optics, Light and Lasers} (Wiley-VCH Verlag GmbH and Co., ISBN 978-3-527-40628-9) p 404
\bibitem{Ghosh} Dey  S, Ray B, Ghosh P N,  Cartaleva S, Slavov D 2015 {\it Opt. Commun.} {\bf 356} 378
\bibitem{Moi}Sargsyan A, Sarkisyan D, Papoyan A, Pashayan-Leroy Y, Moroshkin P, Weis A, Khanbekyan A, Mariotti E,  Moi L 2008  \LP {\bf 18} 749
\bibitem{Ler2015} Sargsyan A,  Tonoyan A, Hakhumyan G, Leroy C, Pashayan-Leroy Y, Sarkisyan D 2015
 \EPL {\bf 110} 23001
 \bibitem{Malak} Nikogosyan G, Sarkisyan D, and Malakyan Yu 2004 {\it J. Opt. Technol.} {\bf 71}, 602–607.
\bibitem{Shore} Shore BW 1990 {\em The Theory of Coherent Atomic Excitation} (Wiley, New York).
\bibitem{DucJOSAB} Dutier G, Saltiel S, Bloch D, and Ducloy M 2003 {\it JOSA }B {\bf 20}, 793–800.





\end{thebibliography}
\end{document}